\documentclass[12pt]{article}
\begin{document}
\noindent
{\large\bf A remark on the choice of stochastic transition rates in driven nonequilibrium systems}

\bigskip\noindent
Hal Tasaki\footnote{
Department of Physics, Gakushuin University,
Mejiro, Toshima-ku, Tokyo 171-8588, Japan
}

\begin{abstract}
We study nonequilibrium steady states of the driven lattice gas with two particles, using the most general stochastic transition rules that satisfy the local detailed balance condition.
We observe that i)~the universal $1/r^d$ long range correlation may be found already in the two-particle models, but ii)~the magnitude (or even the existence/absence) of the long range correlation depends crucially on the rule for transition rates.
The latter is in stark contrast with equilibrium dynamics, where all rules give essentially the same results provided that the detailed balance condition is satisfied.
\end{abstract}

Stochastic processes with discrete state space are often studied as convenient idealized models of various physical systems.
As for time evolution at or close to equilibrium, it has been established \cite{Ons,BL} that transition rates must satisfy the {\em detailed balance condition}\/ in order for the system to obey a macroscopic symmetry known as reciprocity.
But no such criteria are known for systems far from equilibrium.
A standard convention (see, for example, \cite{KLS,S,SZ}) is to take transition rates satisfying the {\em local detailed balance condition}\/, which is a direct generalization of the detailed balance condition.
It seems that an implicit assumption has been that the specific choice of transition rates does not affect physics in a serious manner.
We here show that {\em this is far from the case in driven nonequilibrium systems}\/ .

We here study the driven lattice gas \cite{KLS,S,SZ}, which is one of the standard models of nonequilibrium systems driven by an external force.
We use the most general transition rates satisfying the local detailed balance condition, and obtain the exact steady states of the models when there are only {\em two particles}\/.
We first drive a simple condition under which the steady state is the same as the equilibrium state.
The condition is satisfied in a large class of rules, but not in some of the standard ones including the Metropolis and the heat bath rules.
Then we turn to  transition rates which do not satisfy the condition, and investigate nonequilibrium corrections to the steady states.
We find that, in the dimensions $d\ge2$, {\em the two-particle models may exhibit the universal $1/r^d$ long range correlation}, which is often regarded as an essential feature of nonequilibrium steady states.
Moreover {\em the magnitude of the long range correlation depends crucially on the choice of transition rates}\/: it is extremely large in the Metropolis rule while it is absent in the exponential rule.
Although there were some remarks \cite{previous} about rule dependence in the driven lattice gas, they were mainly quantitative and did not show the sharp implication on the long range correlation.
As for the one-dimensional lattice gas driven by boundary conditions, it was shown recently \cite{vWR} that the model with the exponential rule has a long-range correlation while that with the zero-range rule has no correlation.

\paragraph*{Model:}
Let $\Lambda=\{-(L-1)/2,\ldots,(L-1)/2\}^d\subset{\bf Z}^d$ be the $d$-dimensional $L\times\cdots\times L$ hypercubic lattice where $d\ge2$.
We impose periodic boundary conditions.
There are two identical particles on $\Lambda$, whose positions are denoted by $x\in\Lambda$ and $y\in\Lambda$.
The configuration of the model is given by a pair $(x,y)\in\Lambda\times\Lambda$, where we do not exclude the possibility of $x=y$.
(See the``standard DLG" section for the treatment of hard core repulsion.)
The two particles interact with each other via a potential $V(x-y)=V(y-x)$, and are driven by a constant (electric) field ${\bf E}=(E,0,\ldots,0)$.

We denote by $\cal U$ the set of $2d$ unit vectors.
Let us define a stochastic dynamics by specifying transition rates for any $(x,y)$ and $\delta\in\cal U$ as
\begin{eqnarray}
&&c[(x,y)\to(x+\delta,y)]=\Phi(V(x-y),V(x+\delta-y);{\bf E}\cdot\delta),
\nonumber\\
&&c[(x,y)\to(x,y+\delta)]=\Phi(V(x-y),V(x-y-\delta);{\bf E}\cdot\delta),
\label{e:cP}
\end{eqnarray}
where $\Phi(v,v';E)$ is a function (of three variables) satisfying the local detailed balance condition
\begin{equation}
\Phi(v,v';E)=e^{v-v'+E}\,\Phi(v',v;-E).
\label{e:LDB}
\end{equation}
This reduces to the standard detailed balance condition if $E=0$.
In many cases \cite{KLS} one uses $\Phi$ which can be written as $\Phi(v,v';E)=\phi(-v+v'-E)$ with a function $\phi(h)$ (of a single variable) satisfying $\phi(h)=e^{-h}\,\phi(-h)$.
Among the standard choices are i)~the {\em exponential rule}\/ with $\phi(h)=e^{-h/2}$, ii)~the {\em heat bath (or Kawasaki) rule}\/ with $\phi(h)=2/(1+e^h)$, and iii)~the {\em Metropolis rule}\/ with $\phi(h)=1$ if $h\le0$ and $\phi(h)=e^{-h}$ if $h\ge0$.

We are interested in the properties of the steady state distribution (stationary measure) $p_{x,y}$, which is the unique solution of 
\begin{eqnarray}
&&\sum_{\delta\in{\cal U}}
\Bigl\{-p_{x,y}\,c[(x,y)\to(x+\delta,y)]-p_{x,y}\,c[(x,y)\to(x,y+\delta)]
\nonumber\\
&&
+p_{x+\delta,y}\,c[(x+\delta,y)\to(x,y)]
+p_{x,y+\delta}\,c[(x,y+\delta)\to(x,y)]\Bigr\}
=0,
\label{e:pxy}
\end{eqnarray}
for any $(x,y)\in\Lambda\times\Lambda$.
By using the translation invariance, one can write the steady state distribution as $p_{x,y}=e^{-V(x-y)}\,\tilde{p}_{x-y}$.
Note that we have extracted the equilibrium distribution.
We find from (\ref{e:pxy}) that $\tilde{p}_z$ is determined by
\begin{equation}
\sum_{\delta\in{\cal U}}
[-\tilde{p}_{z}\,\tilde{c}(z\to z+\delta)
+\tilde{p}_{z+\delta}\,\tilde{c}(z+\delta\to z)]=0,
\label{e:pt}
\end{equation}
for any $z\in\Lambda$, where
\begin{equation}
\tilde{c}(z\to z+\delta)=
e^{-V(z)}
\sum_{\sigma=\pm1}\Phi(V(z),V(z+\delta);({\bf E}\cdot\delta)\sigma),
\label{e:ct}
\end{equation}
are the effective transition rates.

\paragraph*{Exponential condition:}
If the effective transition rates satisfy
\begin{equation}
\tilde{c}(z\to z+\delta)=\tilde{c}(z+\delta\to z),
\label{e:cc}
\end{equation}
for any $z\in\Lambda$ and $\delta\in\cal U$, then the solution of (\ref{e:pt}) is simply given by $\tilde{p}_z=\rm const$.
This means that the stationary distribution $p_{x,y}$ is the same as the equilibrium distribution.

Let us examine the condition (\ref{e:cc}).
Let $e_1=(1,0,\ldots,0)$, and $\cal U'$ be the set obtained by removing $\pm e_1$ from $\cal U$.
For $\delta\in\cal U'$ (where we have ${\bf E}\cdot\delta=0$) the condition (\ref{e:cc}) is automatically satisfied because of the (local) detailed balance condition (\ref{e:LDB}) with $E=0$.
For $\delta=\pm e_1$, the condition (\ref{e:cc}) reads
$e^{-v}\{\Phi(v,v';E)+\Phi(v,v';-E)\}=e^{-v'}\{\Phi(v',v;E)+\Phi(v',v;-E)\}$,
which, using (\ref{e:LDB}), can be written as
\begin{equation}
\Phi(v,v';E)=e^E\,\Phi(v,v';-E),
\label{e:exc}
\end{equation}
for any $v$, $v'$, and $E$.
This {\em exponential condition} is satisfied if $\Phi(v,v';E)=e^{E/2}\,\Psi(v,v')$ with any equilibrium transition rate $\Psi(v,v')$ satisfying the detailed balance $\Psi(v,v')=e^{v-v'}\,\Psi(v',v)$.
Among the three standard rules we mentioned, only the exponential rule satisfies the exponential condition (\ref{e:exc}).

\paragraph*{Long-range correlation:}
We now turn to the rules where the exponential condition (\ref{e:exc}) is not satisfied, and investigate nonequilibrium corrections.
For simplicity we only consider the case with on-site interaction described by the potential such that $V(o)=v\ne0$ and $V(z)=0$ if $z\ne o$, where $o=(0,\ldots,0)\in\Lambda$ is the origin.
The behavior of nonequilibrium corrections is essentially the same for more complicated interactions.

Since $v(z)=0$ for $z\ne o$, one has $\tilde{c}(z\to z\pm e_1)=s$ for any $z\in\Lambda$ unless $z$ or $z\pm e_1$ is $o$.
Similarly for any $\delta\in\cal U'$, one has  $\tilde{c}(z\to z+\delta)=t$ unless $z$ or $z+\delta$ is $o$.
The only transition rates which are not equal to $s$ or $t$ are $\tilde{c}(o\to\pm e_1)=s'$, $\tilde{c}(\pm e_1\to o)=s''$, and $\tilde{c}(o\to\delta)=\tilde{c}(\delta\to o)=t'$ where $\delta\in\cal U'$.
We assume $s'\ne s''$ since the exponential condition (\ref{e:exc}) is satisfied if $s'=s''$.

Let us write the condition (\ref{e:pt}) for stationarity as $\sum_{z'\in\Lambda}(T_{z,z'}+D_{z,z'})\,\tilde{p}_{z'}=0$ for any $z\in\Lambda$.
Here we defined the hopping matrix $(T_{z,z'})_{z,z'\in\Lambda}$ by $T_{z,z}=-2s-2(d-1)t$, $T_{z,z\pm e_1}=s$, $T_{z,z+\delta}=t$ for all $z\in\Lambda$ and $\delta\in\cal U'$, and $T_{z,z'}=0$ otherwise.
We also defined $(D_{z,z'})_{z,z'\in\Lambda}$ by $D_{o,o}=2(s-s')+2(d-1)(t-t')$, $D_{e_1,e_1}=D_{-e_1,-e_1}=s-s''$, $D_{\delta,\delta}=t-t'$, $D_{o,\pm e_1}=s''-s$, $D_{\pm e_1,o}=s'-s$, $D_{o,\delta}=D_{\delta,o}=t'-t$ for all $\delta\in\cal U'$, and $D_{z,z'}=0$ otherwise.

We write the (unnormalized) steady state distribution as $\tilde{p}_z=1+\psi_z$ and study the behavior of the nonequilibrium correction $\psi_z$.
Since $\sum_{z'}T_{z,z'}=0$, the equation that determines $\psi_z$ is
\begin{equation}
\sum_{z'\in\Lambda}(T_{z,z'}+D_{z,z'})\,\psi_{z'}=-q_z,
\label{e:TDp}
\end{equation}
for any $z\in\Lambda$, where the ``charge'' is defined by $q_z=\sum_{z'\in\Lambda}D_{z,z'}$.
Here it is found that $q_o=2Q$, $q_{\pm e_1}=-Q$, and $q_z=0$ otherwise, where $Q=s''-s'\ne0$.

Note that one has $\psi_z=0$ if $v=0$ or $E=0$.
Therefore, in the lowest order of perturbation in $v$ and $E$, the nonequilibrium correction $\psi_z$ is determined by the standard lattice Poisson equation $\sum_{z'}T_{z,z'}\,\psi_{z'}=-q_z$.
Since the charge $q_z$ describes a quadrupole with the magnitude $Q$, the asymptotic long-distance behavior of $\psi_z$ for $d\ge2$ is readily obtained as
\begin{eqnarray}
\psi_z&\simeq&c\,Q\,s\,\frac{\partial^2}{\partial{z_1}^2}
\{(\frac{{z_1}^2}{s}+\sum_{j=2}^d\frac{{z_j}^2}{t})^{1-(d/2)}\}
\nonumber\\
&\simeq&c'\,Q\,
\frac{(d-1)({z_1}^2/s)-\sum_{j=2}^d({z_j}^2/t)}
{\{({z_1}^2/s)+\sum_{j=2}^d({z_j}^2/t)\}^{1+(d/2)}},
\label{e:psi}
\end{eqnarray}
where $c$ and $c'$ are constants which depend only on the dimension \cite{2d}.
This is nothing but the universal $1/r^d$ decay found in the driven lattice gas and other related models \cite{ZWLV,GMLS}.

It is now apparent that the most important quantity for determining the nonequilibrium correction is the magnitude $Q=\tilde{c}(\pm e_1\to o)-\tilde{c}(o\to\pm e_1)$ of the quadrupole.
By using (\ref{e:ct}) and the local detailed balance condition (\ref{e:LDB}), we get
\begin{equation}
Q=(1-e^{-E})\,\Phi(0,v;E)-(e^E-1)\,\Phi(0,v;-E).
\label{e:Q}
\end{equation}
We of course have $Q=0$ for the exponential rule.
For the heat bath rule, $Q\simeq vE^2/2$ in the lowest order.
As for the Metropolis rule, we have $Q\simeq E^2$ if $|v|\ge|E|$ and $Q\simeq vE$ if $|v|\le|E|$, again in the lowest order.

\paragraph*{Exact solution:}
To see that the power law decay is {\em not}\/ an artifact of the perturbation, we show how one can get the exact solution of (\ref{e:TDp}).
Our strategy is to define the effective charge $\tilde{q}_z$ which are nonvanishing only for $z\in\{o\}\cup\cal U$ so that the exact solution $\psi_z$ of  (\ref{e:TDp}) also satisfies the Poisson equation $\sum_{z'}T_{z,z'}\psi_{z'}=-\tilde{q}_z$.
This is always possible since (\ref{e:TDp}) reduces to  $\sum_{z'}T_{z,z'}\psi_{z'}=0$ for any $z\not\in\{o\}\cup\cal U$, and there are enough degrees of freedom in $\tilde{q}_z$ to fit the solution.
Note that periodic boundary conditions and the (lattice version of) Gauss' law implies $\sum_z\tilde{q}_z=0$.
Then, by examining the symmetry of the model, we find that $\tilde{q}_z$ can be written with some $Q_1$ and $Q_2$ as $\tilde{q}_o=2Q_1+2(d-1)Q_2$, $\tilde{q}_{\pm e_1}=-Q_1$, and $\tilde{q}_\delta=-Q_2$ for $\delta\in\cal U'$.
Therefore the asymptotic behavior (\ref{e:psi}) is still valid if we replace $Q$ with $Q_1-Q_2$.

Note that $\psi_z$ can  be expressed as a linear combination $\psi_z=\sum_{z'}\tilde{q}_{z'}\,G(z-z')$, where $G(z)$ is the solution of $\sum_{z'}T_{z,z'}G(z')=-\bar{\delta}(z)$ where $\bar{\delta}_z=\delta_{z,o}-L^{-d}$.
By substituting the Poisson equation $\sum_{z'}T_{z,z'}\psi_{z'}=-\tilde{q}_z$ into (\ref{e:TDp}), we get $-\tilde{q}_z+\sum_{z'}D_{z,z'}\psi_{z'}=-q_z$.
This can be regarded as simultaneous linear equations for determining $Q_1$ and $Q_2$.
They can be solved, and one gets lengthy formulae (which we do not write down here) that involve $G(z)$ as well as $Q$, $s$, $s'$, $s''$, $t$, and $t'$ .
In the lowest order,  one of course finds $Q_1\simeq Q$ and $Q_2\simeq0$.

\paragraph*{Standard DLG:}
Finally we briefly discuss the case of the ``standard'' driven lattice gas \cite{KLS,S,SZ} with hard core repulsion and nearest neighbor interaction.
We restrict two-particle configuration $(x,y)$ to those satisfying $x\ne y$, and set $V(x-y)=-J$ if $|x-y|=1$ and $V(x-y)=0$ otherwise.
The analysis is essentially the same as the soft core case, and one again gets (\ref{e:exc}) as the condition for the absence of nonequilibrium corrections.
We also get the long range correlation (\ref{e:psi}), where the ``charge'' $Q$ should be ``renormalized'' \cite{LT}.
But the basic behaviors are the same, and one has $Q\simeq -{\rm const.}JE^2$ in the heat bath rule, and $Q\simeq{\rm const.}E^2$ if $|J|\ge|E|$ and $Q\simeq-{\rm const.}JE$ if $|J|\le|E|$ in the Metropolis rule.

\paragraph*{Discussions:}
The universal $1/r^d$ long range correlation is often referred to as one of the essential features of nonequilibrium steady states in driven systems with a conservation law.
The standard theoretical expression in terms of the fluctuating hydrodynamics \cite{S, SZ, GMLS} may suggest that the long range correlation is a hydrodynamic or a many-body phenomenon.
We have shown, however, that the driven lattice gas with only two particles may exhibit the same long range correlation, suggesting that it is essentially a two-body phenomenon.
It is quite likely that long range correlation in the two-particle model carries over to a many-body system.
This is also suggested by a recent systematic expansion for the steady state of driven lattice gases \cite{LT}.

Perhaps more importantly, we have observed that the magnitude (or even the existence/absence) of the long range correlation depends crucially on the rule of transition rates that one uses.
It is vanishing \cite{vanish} in a large class of models with the exponential condition (\ref{e:exc}), while it is fairly large (probably, too large) and may appear in the first order of $E$ in models with the Metropolis rule.
The disagreement is too sever to be regarded as a minor quantitative difference.
We should probably start seriously examining which rule (if any) provides us with truly physical descriptions of nonequilibrium steady sates.
We stress, however,  that it is impossible to decide which rule is ``realistic'' by purely theoretical arguments, since we still do not know of any fundamental principles that determine nonequilibrium steady states.
It would be desirable if one can find (just as in the case of equilibrium dynamics) some macroscopic constraints from a phenomenological point of view \cite{SST} which enable one to determine microscopic dynamics \cite{Evans}.

\bigskip
It is a pleasure to thank Joel Lebowitz, Raphael Lefevere,  Elliott Lieb, Shin-ichi Sasa, Herbert Spohn for useful discussions.


\begin{thebibliography}{10}
\bibitem{2d}
If $d=2$, one replaces $(\cdots)^{1-(d/2)}$ with $\log(\cdots)$.
One has no power law decay in $d=1$ since the charges are perfectly screened.
\bibitem{vanish}
These models {\em do}\/ have nonequilibrium long-range corrections when there are three or more particles.
We conjecture, however, that the models with the exponential condition (\ref{e:exc}) is free from the $1/r^d$ long range correlation even in the many-body case.
This point may be crucial since the infinite volume steady state can be a Gibbs state only when the $1/r^d$ correlation is absent.
See \cite{LT}.
\bibitem{Evans}
In R. M. L. Evans, Phys. Rev. Lett. {\bf 92}, 150601-1, it is proposed that one should start from the assumption of maximum entropy.
But we simply do not understand the reason.
\end{thebibliography}
\end{document}